\documentclass{article}

\usepackage{moreverb}

\usepackage[colorlinks]{hyperref}

\usepackage{authblk}
\usepackage[justification=centering]{caption}
\usepackage{amsmath}
\usepackage{comment}
\usepackage{placeins}

\usepackage{listings}
\usepackage{tikz}
\usetikzlibrary{shapes,arrows,shapes.multipart}
\renewcommand{\(}{\left(}
\renewcommand{\)}{\right)}
\definecolor{deepblue}{rgb}{0,0,0.5}
\definecolor{deepred}{rgb}{0.6,0,0}
\definecolor{deepgreen}{rgb}{0,0.5,0}
\DeclareFixedFont{\ttb}{T1}{txtt}{bx}{n}{10} 
\DeclareFixedFont{\ttm}{T1}{ttfamily}{c}{n}{9}  
\newcommand\pythonstyleinline{\lstset{
	language=Python, 
	basicstyle = \ttfamily\small,
	basewidth  = 0.48em,
  	otherkeywords={self}, 
	keywordstyle=\bf\color{deepblue}, 
	deletendkeywords={min, object, property},
	emph={MyClass,__init__},
	emphstyle=\bf\color{deepred}, 
	stringstyle=\color{deepgreen}, 
	frame=tb, 
	showstringspaces=false }}
\newcommand\pythonstyle{\lstset{
	language=Python,
	basicstyle = \ttfamily\small,
	basewidth  = 0.48em,
	otherkeywords={self},             
	keywordstyle=\bf\color{deepblue},
	deletendkeywords={min, object, property},
	emph={MyClass,__init__},          
	emphstyle=\bf\color{deepred},    
	stringstyle=\color{deepgreen},
	showstringspaces=false }}
\newcommand\codestyleinline{\lstset{
	language=Python,
	basicstyle = \ttfamily\small,
	basewidth  = 0.48em,
	otherkeywords={},
	deletendkeywords={min, object, property},             
	emph={MyClass,__init__},          
	emphstyle=\bf\color{deepred},    
	stringstyle=\color{deepgreen},
	frame=tb,                        
	showstringspaces=false }}
\newcommand\pythoninline[1]{{\pythonstyleinline\lstinline!#1!}}
\newcommand\codeinline[1]{{\codestyleinline\lstinline!#1!}}
\lstnewenvironment{python}[1][xleftmargin=5mm]
{
\pythonstyle
\lstset{#1}
}{}

\begin{document}

\date{}

\title{Numerical implementation of the multicomponent potential theory of adsorption in Python using the NIST Refprop database}

\author{Rapha\"el Gervais Lavoie\footnote{Email: \url{raphael.gervaislavoie@uqtr.ca}}}
\author{Mathieu Ouellet}
\author{Jean Hamelin}
\author{Pierre B\'enard}

\affil{Institut de recherche sur l'hydrog\`ene, Universit\'e du Qu\'ebec \`a Trois-Rivi\`eres, \break
D\'epartement de chimie, biochimie et physique, Universit\'e du Qu\'ebec \`a Trois-Rivi\`eres.}

\maketitle

\begin{abstract}
In this paper, we present a detailed numerical implementation of the multicomponent potential theory of adsorption which is among the most accurate gas mixtures adsorption models. The implementation uses the NIST Refprop database to describe fluid properties and applies to pure gases and mixtures in both subcritical and supercritical regimes. The limitations of the model and the issues encountered with its implementation are discussed. The adsorption isotherms of CH$_4\,/\,$CO$_2$ mixture are modeled and parameterized as implementation examples. 
\end{abstract}


\vspace{-6pt}

\section{Introduction}
\vspace{-2pt}
Gas adsorption in porous material is relevant to a wide variety of industrial and scientific applications, ranging from gas purification and separation to gas storage in stationary or mobile applications. While adsorption experiments of pure gases can be performed readily, experiments on gas mixtures can be challenging, expensive and time-consuming \cite{talu1998needs}. This situation motivates the utilization of theoretical models to predict the behavior of gas mixtures in the presence of adsorbent.

Several approaches have been proposed to model the adsorption isotherms of both pure gases and mixtures \cite{henry1803experiments, freundlich1906over, langmuir1918adsorption, brunauer1938adsorption, temkin1940recent, sips1948structure, sips1950structure, kisliuk1957sticking, toth1971state, dubinin1971development, ruthven1976sorption, suwanayuen1980gas, doong1988simple, myers2002thermodynamics}.  In this paper, we will concentrate on the potential theory of adsorption (PTA), a two-parameter thermodynamic model developed by Shapiro and Stenby \cite{Shapiro1998}, based on the pore filling approach of Polanyi's  theory of adsorption \cite{Polanyi1963}.
For gas mixtures adsorption, the PTA model generalized into the multicomponent potential theory of adsorption (MPTA), which can deliver great performance, especially when considering disordered adsorbents such as activated carbon (an overview of mixture adsorption models is presented in Ref. \cite{Dundar2012}). For a $N$ components gas mixture, the simplest MPTA model required the adjustment of only $N+1$ parameters, obtained from pure components adsorption isotherms.

The objective of this paper is to present a detailed numerical implementation of the MPTA model. In this spirit, we will write down explicit lines of code, written in Python3, an increasingly popular, easy to follow, non-compiled and non-proprietary programming language. We will also point out pitfalls and subtleties encounter, and show how to identify and deal with phase transitions. 

Definitions and symbols used in this paper are described in Table \ref{t0}.

\begin{table}[!htb]
\centering
\captionof{table}{Variables definition.}\label{t0}
\begin{tabular}{|l|c|l|}
\hline
Variables & Type & Definition \\
\hline
\hline
$\mu_B\ $, \pythoninline{chem()}		& function		& Chemical potential of the bulk phase (J/mol)\\
$\mu_{Ad}\ $, \pythoninline{chem()}	& function		& Chemical potential of the adsorbed phase (J/mol)\\
$\varepsilon\ $, \pythoninline{DRA()}	& function		& Adsorbent surface potential (J/mol)\\
$\rho_B\ $, \pythoninline{db}		& float		& Fluid density in the bulk phase (mol/L)\\
$\rho_{Ad}\ $, \pythoninline{d_z}	& float		& Fluid density in the adsorbed phase (mol/L)\\
$\varepsilon_0\ $, \pythoninline{eps0}& float, array	& Characteristic energy of adsorption (J/mol)\\
$z\ $, \pythoninline{z}				& float		& Microporous volume (cm$^3$/g)\\
$z_0\ $, \pythoninline{z0}			& float		& Limiting micropore volume (cm$^3$/g)\\
$\beta$						& float		& Heterogeneity parameter (fix to 2 in this work)\\
$N_{ex}\ $, \pythoninline{N_ex_}	& function		& Excess (Gibbs) adsorption (mol/Kg) \\
\pythoninline{dataP}				& array		& Experimental values of pressure (KPa) \\
\pythoninline{dataD}				& array		& Corresponding values of fluid density (mol/L) \\
\pythoninline{dataAd}			& array		& Experimental values of excess adsorption (mol/Kg) \\
$x_B\ $, \pythoninline{xB}			& array		& Bulk phase molar fraction array \\
$x_{Ad}\ $, \pythoninline{x_z}		& array 		& Adsorbed phase molar fraction array \\
\pythoninline{x_z_A} , \pythoninline{x_z_B}		& float		& Adsorbed phase molar fraction of components \\
\pythoninline{d_max}			& function		& Maximal fluid density allowed by the Refprop (mol/L)\\
\pythoninline{d_vap}				& float, array	& Fluid vapor density at dew point (mol/L)\\
\pythoninline{d_liq}				& float, array	& Fluid liquid density at dew point (mol/L)\\
\pythoninline{X}				& array		& Array containing pure gases arrays \\
\pythoninline{d0}				& float		& Initial guess for fluid density (mol/L) \\
\pythoninline{x0}				& array		& Initial guess for mixture molar fraction array \\ 
\pythoninline{T}				& float		& Fluid temperature (K) \\
\hline
\end{tabular}
\end{table}

\vspace{-6pt}

\section{The multicomponent potential theory of adsorption model}
\vspace{-2pt}

When dealing with adsorption, it is useful to define two regions for the adsorbate, the \emph{bulk phase} and the \emph{adsorbed phase}. The \emph{bulk phase} defines a region far from the adsorbent surface where the fluid \emph{is not significantly affected} by the presence of the adsorbent material. Conversely, the \emph{adsorbed phase} defines a region near the adsorbent surface where the fluid \emph{is significantly affected} by the adsorbent material. The thermodynamic properties of the \emph{bulk phase} are described by the \emph{equation of state} (EoS) of the fluid. 

\subsection{Pure gas}

The main assumption of the MPTA model is to suppose that the fluid\nobreakdash--surface interaction is completely described by a local potential field $\varepsilon$, generated by the surface. The potential $\varepsilon$ is attractive near the surface of the adsorbent and tends to vanish far from the surface. 
The constitutive equation of the MPTA model is given by Refs. \cite{myers2002thermodynamics, Shapiro1998}
\begin{align}
\mu_B\(T,\rho_B\) = \mu_{Ad}\(T,\rho_{Ad}\) - \varepsilon, \label{mu}
\end{align}
where $\mu_{B}$ and $\rho_B$ are the chemical potential and the fluid density in the bulk phase, while $\mu_{Ad}$ and $\rho_{Ad}$ are the locals chemical potential and fluid density in the adsorbed phase. The bulk phase properties are assumed to be constant while the adsorbed phase properties vary with position \cite{Shapiro1998}. Hence, the density $\rho_{Ad}$ is usually very high inside pores, close to the adsorbent, and decrease to reach $\rho_B$ far from the adsorbent. Given Eq. \eqref{mu}, the local thermodynamic properties in the adsorbed phase are uniquely determined from properties of the bulk phase and the interaction potential $\varepsilon$. Since we consider only equilibrium situations, the temperature is constant.

Along with Eq. \eqref{mu}, the additional information needed is an explicit form for the potential $\varepsilon$. We use the Dubinin--Radushkevich--Astakhov (DRA) equation \cite{Dubinin1989}, a two-parameter ($\varepsilon_0$ and $z_0$), semi-empirical potential originates from the theory of volume filling micropores \cite{hutson1997theoretical, dubinin1985generalization} and which can represent a wide range of energy distributions
\begin{align}
\varepsilon(z) = \varepsilon_0 \( \ln{\frac{z_0}{z}} \)^{\frac{1}{\beta}}. \label{DRA}
\end{align}
Here, $\varepsilon_0$ is the \emph{characteristic energy of adsorption} of the adsorbate on the adsorbent,  a parameter representing the magnitude of the fluid-solid interaction. The variable $z$ has units of a volume and $z_0$ is the \emph{limiting micropore volume} of the adsorbent which is the volume of the pore that significantly contributes to the adsorption.
 The ratio $z/z_0$ represents the fraction of the microporous volume associated to an energy $\varepsilon(z)$. $\beta$ is a measure of the heterogeneity of the adsorbent and is usually set to 2 for activated carbon, but can eventually become a fitting parameter (in this paper, we will use $\beta=2$). Under certain conditions,  the variable $z$ can be seen as the \emph{distance} from the surface of the adsorbent. This is the case when the microporous structure of the adsorbent exhibit a planar symmetry as in slit-pore adsorbents for example. In this paper, we will use this more intuitive approach and interpret $z$ as the distance from the surface of the adsorbent as in Ref. \cite{Shapiro1998}. 

The presence of the divergence in Eq. \eqref{DRA} as $z\to0$ is handled, as shown bellow, by setting a cut-off value for the chemical potential in the numerical implementation.

Eq. \eqref{mu} now need to be inverted to obtain the local density $\rho_{Ad}(z)$ from the chemical potentials. Then, the (Gibbs) excess adsorption $N_{ex}$ (which is what's experimentally measured), is calculated by
\begin{align}
N_{ex}(\rho_B) = \int_0^{z_0} \(\rho_{Ad}(z) - \rho_B\)dz.\label{Nex}
\end{align}

We should emphasize a crucial point here: even if $\varepsilon_0$ and $z_0$ have well-defined physical meaning, these two parameters are better used as fitting parameters in the model. Values for the characteristic energy of adsorption ($\varepsilon_0$) for physisorption on microporous adsorbents range from 5--15 kJ/mol while typical limiting micropore volume ($z_0$) range from 0.1 to 2 cm$^3$/g. 

To fit the model to experimental datasets, initial guess of $\varepsilon_0$ and $z_0$ are made. The differences between the calculated excess adsorption ($N_{ex}(\rho_B)$) and the experimental data at various pressures are then simultaneously minimized.  The result of the minimization between calculated and experimental values of excess adsorption translates into optimal values for $\varepsilon_0$ and $z_0$.

In the literature, many authors used the fugacity $f$ instead of the chemical potential. In that case, Eq. \eqref{mu} takes the form
\begin{align}
f_{Ad}\(T,\rho_{Ad}(z)\)= f_B\(T,\rho_B\)\exp\(\frac{\varepsilon(z)}{RT}\).
\end{align}
Fugacity being essentially the exponential of the chemical potential, we choose to work with the later. Also, it is worth noting that when working with the chemical potential, the logarithmic singularity of Eq. \eqref{DRA} is integrable, which is not the case when working with fugacity.

\subsection{Mixtures}\label{Mixtures}

For gas mixtures, the situation is more challenging, particularly for the description of the local chemical potential in the adsorbed phase. The molar fraction of the component $i$ will be noted $x^i$ such that $\sum_i^M x^i=1$ where $M$ is the number of pure components in the mixture. While the molar fraction of the bulk phase can be well known from experimental measurements (gas chromatography for instance), the molar fraction in the adsorbed phase is unknown and may vary substantially from the bulk phase. In gas mixtures, each gas component interacts with the surface through is own potential
\begin{align}
\varepsilon^i(z) = \varepsilon^i_0 \( \ln{\frac{z_0}{z}} \)^{\frac{1}{\beta}},
\end{align}
where $i=1\dots M$. Due to the selectivity of adsorbent material, the adsorption of one component will be favored, affecting the local molar fraction of the mixture in the adsorbed phase.

The introduction of the molar fractions adds unknown variables to the problem. Instead of simply inverting Eq. \eqref{mu} to find $\rho_{Ad}(z)$, we need to solve simultaneously the following system of equations \cite{monsalvo2009study}
\begin{align}
\mu_B^i(\rho_B,x_B^i) + \varepsilon^i(z) - \mu_{Ad}^i(\rho_{Ad}(z),x_{Ad}^i(z)) &= 0,\label{chempot}
\end{align}
for $x_{Ad}^i(z)$ and $\rho_{Ad}(z)$, where the superscript $i$ refer to the gas component of the mixture. From now on, the chemical potential also depend on molar fraction $x^i$ of gas components ($x_B^i$ is the molar fraction of the component $i$ of the bulk phase at equilibrium, {\it i.e.} when the adsorption process is complete). In Eq. \eqref{chempot}, $\rho_B$, $x^i_B$ and $\varepsilon^i(z)$ are known and $\rho_{Ad}(z)$ and $x^i_{Ad}(z)$ are the unknowns. The dimension of the system of equation to solve is $M$. The set of $x_{Ad}^i(z)$ represent $M-1$ independent variables (because $\sum_i^M x_{Ad}^i=1$ decrease by 1 the number of independent variables) and there is one more independent variable $\rho_{Ad}(z)$ which, of course, is the same for all gas components. 

Once the $x_{Ad}^i(z)$ and $\rho_{Ad}(z)$ are determined, the next step is to calculate the excess adsorption for each component:
\begin{align}
N_{ex}^i(\rho_B) = \int_0^{z_0}\( \rho_{Ad}(z)x_{Ad}^i(z) - \rho_Bx_B^i \)dz.
\end{align}

Finally, the total amount of excess gas adsorbed is simply given by
\begin{align}
N_{ex}(\rho_B) = \sum_{i=1}^M N_{ex}^i(\rho_B).
\end{align}

As for the pure gases, to adjust the model to experience, initial values of $\varepsilon_0^i$ and a common $z_0$ must be postulated. Because $z_0$ is considered to be an adsorbent property (the limiting micropore volume), we use a single value for all gas components. The excess adsorption is calculated for each gas component from the postulated initial values. Then, the differences between the calculated and experimental pure excess adsorption is minimized simultaneously for all experimental bulk phase densities and all gas components.
This finally gives the optimal values of parameters  $\varepsilon_0^i$ and a common $z_0$. 

A key point here is that even for mixtures, the adjustment of the model to experimental data is made using \emph{pure gas adsorption isotherms}, not mixture data \cite{monsalvo2009study}. Once the parameters are adjusted from pure gas isotherms, then gas mixture adsorption can be simulated. It is important, however, to understand that $x_B$ \emph{should be the bulk phase molar fraction at equilibrium which may differ greatly from the initial or feeding molar fraction}. For example, in Ref. \cite{Sudibandriyo2003}, the feeding and equilibrium composition can easily differ by more than 25\%. Then, using initial instead of equilibrium molar fractions could introduce a significant error in the model predictions. Consequently, one should be careful when using the MPTA model on data with unknown bulk phase equilibrium molar fractions.

\subsection{Equation of state}

Up to now, there is one thing that we left aside: the way we will evaluate the chemical potentials $\mu^i_B$ and $\mu^i_{Ad}$. 
In order to accurately describe the thermodynamic properties of the fluid, one should use an accurate equation of state (EoS). A well-known EoS commonly used for this kind of purpose is the Soave\nobreakdash--Redlich\nobreakdash--Kwong (SRK) EoS \cite{Redlich1949,Soave1972}. However, in this paper, we choose to use the NIST\nobreakdash--REFPROP (Refprop) database \cite{Refprop} which give presumably the most accurate results. For the purpose of this paper, the underlying references used by the Refprop are Refs. \cite{setzmann1991new, span2000reference, span1996new}. Credit goes to Refs. \cite{Thelen, Wernick} for the Python integration of Refprop.

\subsection{Performance analysis}

To quantify the precision of the model, we will use the absolute average deviations (AAD)  between experimental and fitted data for excess adsorption:
\begin{align}
ADD(N)=\frac{100}{k}\sum_{j=1}^k\bigg| \frac{N_{ex}^{exp}(\rho_B^j)-N_{ex}^{cal}(\rho_B^j)}{N_{ex}^{exp}(\rho_B^j)} \bigg| \%.
\end{align}
In this equation, $\rho_B^j$ is one of the experimental bulk phase density in the set of $k$ values $\{\rho_B^1, \rho_B^2,\dots,\rho_B^k\}$. $N_{ex}^{exp}(\rho_B^j)$ and $N_{ex}^{cal}(\rho_B^j)$ are the experimental and calculated gas adsorbed density at bulk phase density $\rho_B^j$, respectively.

\section{Numerical implementation}
\vspace{-2pt}

The proposed numerical integration being not very computationally expensive, Python (version 3) \cite{Python1, Python2} might be the best choice of language to obtain a simple, concise and open-source implementation. Indeed, with all the embedded functions and algorithms (principally from SciPy \cite{SciPy} and NumPy \cite{NumPy} packages), most of the numerical operations are easier to do in Python than in other languages. Moreover, the Jupyter (IPython) notebook \cite{IPython} and matplotlib \cite{Matplotlib} package provide great and convenient tools for graphical visualization and analysis. 

For readers unfamiliar with Python, the language is simple enough to serve as pseudocode for other languages as well. Also, Section \ref{python} give a short introduction to Python. Note that some programming convention will be overlooked for space consideration and that the proposed implementation is oriented on simplicity and concision rather than performance.

\subsection{Minimal introduction to Python}\label{python}

We present here some of Python concepts needed for the rest of this paper. Firstly, the required package needs to be loaded. In our case, those are the standard packages \pythoninline{numpy, scipy,} \pythoninline{matplotlib, decimal} and the non-standard package \pythoninline{lmfit} \cite{Lmfit}. The packages are loaded with

\begin{minipage}{\linewidth}
\begin{python}
import numpy, scipy, lmfit, decimal
import matplotlib.pyplot
% matplotlib inline
\end{python}
\end{minipage}
The last two lines are required for graphical visualization directly inside the notebook.

Python support object oriented programing, which means that properties or sub-functions can be call using dotted notation \codeinline{object.property()}. For instance, all the Refprop functions will be called using the syntax \pythoninline{refprop.function_name()}.

Arrays can be created explicitly like \pythoninline{x=[1,2,3,4,5]}, or can be defined generically and values added when required:

\begin{minipage}{\linewidth}
\begin{python}
y = [ ]
y.append(5)
\end{python}
\end{minipage}
Values are extracted with $\cdot$\pythoninline{[i]}, where \pythoninline{i} is the position in the array, starting from 0. In our example, \pythoninline{x[1]=2} and \pythoninline{y[0]=5}. The length of the array \pythoninline{x} is given by \pythoninline{len(x)}, and the array can by printed out explicitly with \pythoninline{print(x)}.

A function $f(x,y,z)=x^2+y^2-\text{e}^z$ can be define by

\begin{minipage}{\linewidth}
\begin{python}
def f(x, y, z):
    function = x**2 + y**2 - numpy.exp(z)
    return function
\end{python}
\end{minipage}
Note that the indentation here is crucial: Python use indentation to structure the code instead of  braces like other languages. The same code with no indentation will produce an error.

To shorten this paper, variables assignment will sometimes be written on a single line

\begin{minipage}{\linewidth}
\begin{python}
x, y, z = 5, 'yes', [1, 2, 3]
\end{python}
\end{minipage}
On this single line, the value 5 is assign to the variable \pythoninline{x}, the string \pythoninline{'yes'} is assign to \pythoninline{y} and the array \pythoninline{[1,2,3]} is assign to \pythoninline{z}.
Let us write

\begin{minipage}{\linewidth}
\begin{python}
x = numpy.arange(0, 50)
y = x**2
\end{python}
\end{minipage}
The first line creates an array \pythoninline{x} containing the values from 0 to 49 while the second line creates an array \pythoninline{y} where each value is the square of the ones in the \pythoninline{x} array. In Python, there is a distinction between a list and an array, but we will overlook it since it is basically a matter of performance. The term \emph{array} will then be used for both lists and arrays. Finally, a simple point plot of \pythoninline{y} as a function of \pythoninline{x} can be made with

\begin{minipage}{\linewidth}
\begin{python} 
fig = matplotlib.pyplot.plot(x, y)
matplotlib.pyplot.show(fig)
\end{python}
\end{minipage}

Refprop Python functions usually return results in the form of a \emph{dictionary data structure}, which associate a \emph{key} (a string) to a \emph{value}. Dictionaries are in fact simply arrays where elements are extracted using a key instead of an index. Then, to extract the numerical value of a variable in a dictionary data structure, we need to add the key associated to that variable in brackets after the function. For example, if the function \pythoninline{f()} return the dictionary

\begin{minipage}{\linewidth}
\begin{python}
f = {'P': 5, 'T': 298, 'SubCrit': True},
\end{python}
\end{minipage}
 the command \pythoninline{f['P']} will return the numerical value 5.

\subsection{Simplest case: Supercritical pure gas}

In this section, we implement the simplest case of a supercritical pure gas far from its critical point. The possibles complications will be added successively in the following sections.

\subsubsection{Evaluation of the excess adsorption.}

First, we need to import the Refprop package, define the installation path and initialize the fluid that we want to work with (methane in our example):

\begin{minipage}{\linewidth}
\begin{python}
import refprop
refprop.setpath(path='/usr/local/lib/refprop')
refprop.setup(u'def', u'METHANE')
\end{python}
\end{minipage}

Next, we need to import experimental dataset of bulk phase pressure and excess adsorption.
In the following, we will use experimental data from Ref. \cite[run 1]{Sudibandriyo2003}.  We can use the function \pythoninline{numpy.loadtxt()} to import data from a text file.
Assuming that the first column in \pythoninline{data.txt} contain the experimental pressure values of the bulk phase (in KPa) and the second column (separated by white spaces) contain the excess gas adsorbed (in mol/Kg), then the code

\begin{minipage}{\linewidth}
\begin{python}
dataP  = numpy.loadtxt('data.txt', dtype='float', usecols=[0])
dataAd = numpy.loadtxt('data.txt', dtype='float', usecols=[1])
\end{python}
\end{minipage}
creates the two arrays \pythoninline{dataP} and \pythoninline{dataAd} for pressure and excess adsorption, respectively. 

Refprop functions usually required gas density instead of pressure. The gas density can be obtained using \pythoninline{refprop.press(T,d,x)}, with \pythoninline{T} the temperature (in K), \pythoninline{d} the density (in mol/L) and \pythoninline{x} the gas molar fraction array (for pure gases, \pythoninline{x=[1]}, for a 25\%\,/\,75\% mixture, \pythoninline{x=[0.25, 0.75]}, etc). The gas molar fraction array will be noted \pythoninline{xB} and \pythoninline{x_z} for the bulk phase and the adsorbed phase (at position \pythoninline{z}) respectively. Of course, for pure gases, \pythoninline{xB=x_z=x=[1]}. It can be noted that we used \pythoninline{d} for the density $\rho$ and \pythoninline{eps0} for $\varepsilon_0$ for convenience. To transform the pressure array \pythoninline{dataP} into a density array \pythoninline{dataD}, we first define a function \pythoninline{pressure} that depend on density as 

\begin{minipage}{\linewidth}
\begin{python}
def pressure(d, T, x, p):
    return refprop.press(T, float(d), x)['p'] - p
\end{python}
\end{minipage}
Then, for each value of pressure in \pythoninline{dataP[i]} ($i=0\dots N-1$, where $N$ is the number of data), we solve the equation \pythoninline{pressure(d,T,xB,p)=0} to find the associated density \pythoninline{d}. We use the function \pythoninline{scipy.optimize.fsolve(f,x0,args=(...))}  for that purpose (which is based on MINPACK's hybrd and hybrj algorithms), but any root finding algorithm would work. Here, \pythoninline{f} is the function to be solved (assuming the form \pythoninline{f=0}) and \pythoninline{x0} is an initial guess for \pythoninline{f}. If the function \pythoninline{f} is multidimensional (as in our case), the parameter \pythoninline{args=(...)} enable the specification of all the variables but the first in \pythoninline{f}. For example, if we want to solve \pythoninline{f(u,v,w)=0} with respect to \pythoninline{u}, using the initial guess \pythoninline{u=1} and for the specific values \pythoninline{v0} and \pythoninline{w0}, we use \pythoninline{scipy.optimize.fsolve(f,1,args=(v0,w0))}.

Because there is a bijection between pressure and density, the specification of \pythoninline{T}, \pythoninline{xB} and \pythoninline{p} in \pythoninline{pressure()} function generate a unique root and then, any reasonable guess for the density will work at this stage, so we take \pythoninline{d0=1}. Explicitly, this give

\begin{minipage}{\linewidth}
\begin{python}
dataD = [ ]
for i in range(0, len(dataP)):
    d = scipy.optimize.fsolve(pressure, 1, args=(T, xB, dataP[i]))
    dataD.append(float(d[0]))
\end{python}
\end{minipage}
Simply state, this code solve the equation \pythoninline{pressure(d,T,xB,p)=0} with respect to \pythoninline{d} for known values of \pythoninline{T}, \pythoninline{xB} and \pythoninline{p}, given by \pythoninline{args=(T,xB,dataP[i])}, starting from an initial guess \pythoninline{d0=1}, and with $i\in(0,\dots,N-1)$. The results are then put in the array  \pythoninline{dataD} in the last line. Note that \pythoninline{fsolve()} return an array containing \emph{numpy.float64} data type here. To get the numerical value only (meaning a standard \emph{float} data type), we have to use \pythoninline{float(d[0])}, which take the first element of the array \pythoninline{d} and ensures a \emph{float} data type. Python float data type correspond to the usual \emph{double precision} (binary64) data type \cite{ieee2008754}.

Next, we need to know the maximum density allowed by the Refprop for the gas considered. We will define a function \pythoninline{d_max()} to lighten the code. This function will take the temperature \pythoninline{T} and the molar fraction array \pythoninline{x} of the mixture (\pythoninline{x=[1]} everywhere for pure gas), pass it to the function \pythoninline{refprop.limitx()} and extract the maximum density allowed: 

\begin{minipage}{\linewidth}
\begin{python}
def d_max(x):
    return refprop.limitx(x, htype='EOS', t=T, D=0, p=0)['Dmax']
\end{python}
\end{minipage}

The Refprop also have an embedded function for the chemical potential \pythoninline{chempot(T,d,x)['u']}. This function returns an array containing the chemical potential of all the component of a gas mixture with molar fraction array given by \pythoninline{x}. For simplicity, let us define a function \pythoninline{chem} as

\begin{minipage}{\linewidth}
\begin{python}
def chem(T, d, x):
    return refprop.chempot(T, float(d), x)['u']
\end{python}
\end{minipage}
In the case of pure gas (or the first component of a gas mixture), \pythoninline{chem(T,d,x)[0]} will return the numerical value of the chemical potential. This notation will be quite useful for mixtures below.

We can also implement the DRA potential Eq. \eqref{DRA} (with $\beta=2$) as

\begin{minipage}{\linewidth}
\begin{python}
def DRA(z0, eps0, z):
    return eps0*numpy.log(z0/z)**(1.0/2.0)
\end{python}
\end{minipage}

From the previous functions, the chemical potential of the fluid at a distance \pythoninline{z} from the surface of the adsorbent can be written as 

\begin{minipage}{\linewidth}
\begin{python}
chem(T, d_z, x_z)[0] = chem(T, dB, xB)[0] + DRA(z0, eps0, z)
\end{python}
\end{minipage}
In this equation, the unknowns variables are \pythoninline{z} and \pythoninline{d_z} (the density in the adsorbed phase at \pythoninline{z}). For the moment, we supposed that the values of \pythoninline{z0} and \pythoninline{eps0} are known. Note that we will use \pythoninline{dB}, \pythoninline{d_z}, \pythoninline{xB} and \pythoninline{x_z} to make it clear that we talk about the bulk or adsorbed phase (remember that \pythoninline{xB=[1]=x_z} for pure gas, but we will leave \pythoninline{xB} and \pythoninline{x_z} here for generality). To  extract the density \pythoninline{d_z}, we need to provide a value for \pythoninline{z} in the range $0\leq z\leq z0$, and solve the following equation

\begin{minipage}{\linewidth}
\begin{python}
0 = chem(T, dB, xB)[0] + DRA(z0, eps0, z) - chem(T, d_z, x_z)[0].
\end{python}
\end{minipage}
Here again, we use \pythoninline{fsolve()} for that purpose, along with an initial guess \pythoninline{d0} for the density (let say \pythoninline{d0=dB}). Explicitly, we can build a density function \pythoninline{d_pure()} as

\begin{minipage}{\linewidth}
\begin{python}
def d_pure(z, T, z0, eps0, dB, xB, x_z):
    def f(d_z):
        y = chem(T, dB, xB)[0] + DRA(z0, eps0, z) - chem(T, d_z, x_z)[0]
        return y
    if chem(T, dB, xB)[0] + DRA(Z0, eps0, z) >= chem(T, d_max(xB), xB)[0]:
        return d_max(xB)
    else:
        return scipy.optimize.fsolve(f, dB)[0]
\end{python}
\end{minipage}
This function return the density of the gas for \pythoninline{z} in the range $0\leq z\leq z0$. On the fifth line, we added a test condition to ensure that the algorithm will not exceed the maximum density allowed by the Refprop. By putting a cut-off on the chemical potential, we in fact assume that the chemical potential is a  monotonous increasing function of the density, which means that $\mu_{max}$ correspond to $\mu(\rho_{max})$. It is this cut-off that takes care of the discontinuity in the DRA potential in Eq. \eqref{DRA} as $z\to0$. The error introduce by this cut-off is negligible over the whole range $0\leq z\leq z_0$. In fact, instead of using the maximal density value \pythoninline{d_max}, the results of the implementation remain unchanged if we use a drastic cut-off value $\rho=0$.

\begin{figure}[!htb]
\centering
\includegraphics[width=12cm]{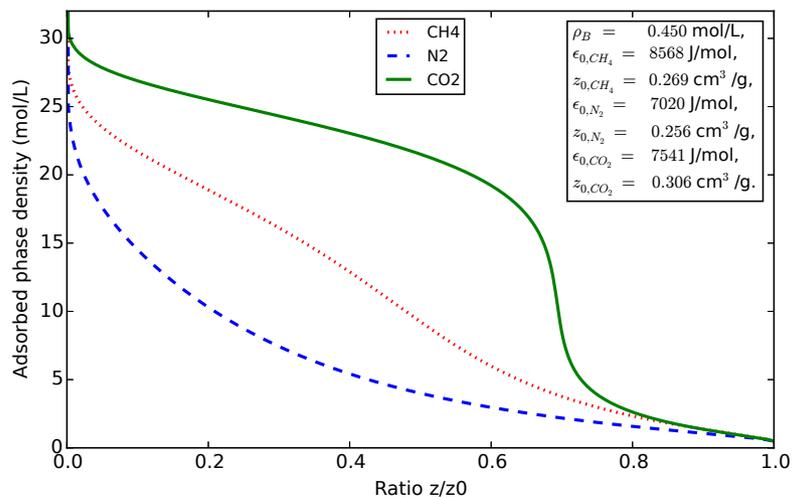}
\caption{\label{DensityProfile-pure}Density profile of pure CH$_4$, N$_2$ and CO$_2$\\  near the surface of Calgon F-400 at 318.2K.}
\end{figure}

Figure \ref{DensityProfile-pure} is an example of the density profile given by \pythoninline{d_pure()} as a  function of the ratio \pythoninline{z/z0} for pure CH$_4$, N$_2$ and CO$_2$ on activated carbon Calgon Filtrasorb 400 at 318.2\,K with a bulk gas density of 0.45 mol/L (corresponding to bulk pressure arround 1.2 MPa). Values of $\varepsilon_0$ and $z_0$ used are the one given in table \ref{t1}. As said before, the experimental data are taken from Ref. \cite[run 1]{Sudibandriyo2003}.

Finally, we can integrate the function \pythoninline{d_pure} on \pythoninline{z} from \pythoninline{0} to \pythoninline{z0} to obtain the adsorbed gas quantity. We use the function \pythoninline{scipy.integrate.quad()} for that purpose:

\begin{minipage}{\linewidth}
\begin{python}
def N_ex_pure(T, z0, eps0, dB, xB):
    y = scipy.integrate.quad(d_pure, 0, z0, args=(T, z0, eps0, dB, xB, xB))
    return y[0] - dB*z0
\end{python}
\end{minipage}
The output of this function is the (Gibbs) excess adsorption (in mol/Kg) considering the temperature, the parameters \pythoninline{z0} and \pythoninline{eps0}, the bulk gas density \pythoninline{dB} (in mol/L) and the molar fraction \pythoninline{xB=[1]}. Note that in the last line, we subtract the gas quantity that would be present in the absence of adsorbent, namely \pythoninline{dB*z0}, in order to obtain the excess adsorption and not the absolute adsorption.

\FloatBarrier
\subsubsection{Building and fitting the model.}

We have shown how the excess adsorption could be evaluated numerically, provided that \pythoninline{z0} and \pythoninline{eps0} are known. In the following, we will show how to obtain those values by fitting the MPTA model to experimental dataset for pure gases.

First of all, minimization algorithms are built to minimize a function. Then, we have to build a function, based on our calculated excess adsorption, which can be minimized with respect to parameters \pythoninline{z0} and \pythoninline{eps0}. To do that, we can define the function \pythoninline{pure_fit} where arguments will be pass by a dictionary data structure this time (as required by the chosen minimization algorithm):

\begin{minipage}{\linewidth}
\begin{python}
def pure_fit(params):
    value = params.valuesdict()
    z0 = value['z0']
    eps0 = value['eps0']
    difference = [ ]
    for i in range(0, len(dataD)):
        difference.append(N_ex_pure(T, z0, eps0, dataD[i], xB) - dataAd[i])
    return difference
\end{python}
\end{minipage}
In words, this function accept the argument \pythoninline{params} which is a dictionary data structure containing \pythoninline{z0} and \pythoninline{eps0}. For each experimental value of density contain in \pythoninline{dataD}, the function takes the difference between the excess adsorption given by \pythoninline{N_ex_pure()}, evaluated with the values \pythoninline{z0} and \pythoninline{eps0}, and the experimental adsorption contain in \pythoninline{dataAd}. Those differences between calculated and experimental adsorption data are put in the array \pythoninline{difference} which have the same dimension as \pythoninline{dataD} and \pythoninline{dataAd}. The upshot is that the function \pythoninline{pure_fit()} do not return a value but rather an array of values. By minimizing the function \pythoninline{pure_fit()}, we, in fact, minimize the array just built and thus, the differences between the calculated and experimental excess adsorption. To do that, we used the function \pythoninline{minimize()} from the package \pythoninline{lmfit} \cite{Lmfit}, which use a Levenberg-Marquardt algorithm for non-linear least-squares minimization. Explicitly, we have

\begin{minipage}{\linewidth}
\begin{python}
params = lmfit.Parameters()
params.add('z0', value=1, min=0, vary=True)
params.add('eps0', value=10000, min=0,  vary=True)
result = lmfit.minimize(pure_fit, params)
print(lmfit.fit_report(result))
\end{python}
\end{minipage}
The function \pythoninline{lmfit.Parameters()} create a dictionary named \pythoninline{params} on the first line. Parameters \pythoninline{z0} and \pythoninline{eps0} are added to the dictionary on the second and third line, specifying the initial and minimal values, and setting them as adjustable parameters with \pythoninline{vary=True}. Then, the minimization takes place on the fourth line on the function \pythoninline{pure_fit()} using parameters \pythoninline{params}. The fitted values of \pythoninline{z0} and \pythoninline{eps0} and their uncertainties are given in the fit report in the last line. The package \pythoninline{lmfit} have the advantage of easily allowing the modification of fitting variables, simply by changing \pythoninline{vary=True} to \pythoninline{vary=False}. To give an idea of the computational time involved, the preceding fit took between 2 min (N$_2$ with 11 data points) to nearly 7 min (CO$_2$ with 13 data points) on an Intel i7-2600 CPU running one thread. 

Finally, from the fitted values \pythoninline{fit_z0} and \pythoninline{fit_eps0} (given by \pythoninline{lmfit.fit_report()}), we can build an array \pythoninline{dataFit} that will contain the calculated excess adsorption to be compared with experimental \pythoninline{dataAd}.

\begin{minipage}{\linewidth}
\begin{python}
dataFit = [ ]
for i in range(0, len(dataD)):
    dataFit.append(N_ex_pure(T, fit_z0, fit_eps0, dataD[i], xB)) 
\end{python}
\end{minipage}
\begin{table}[!htb]
\centering
\caption{Initial and fitted values of the parameters for pure gases.}\label{t1}
\begin{tabular}{|l|c|c|c|c|}
\hline
Parameter & Initial value & fitted value & fit error $\%$\\
\hline
\hline
$z_{0,CH_4}$ ($cm^3/g$)                & 1            & 0.269      & $0.38\%$ \\
$z_{0,N_2}$ ($cm^3/g$)                  & 1             & 0.256      & $0.72\%$ \\
$z_{0,CO_2}$ ($cm^3/g$)               & 1             & 0.306      & $2.16\%$ \\
$\varepsilon_{0,CH_4}$ ($J/mol$)          & $10\,000$     & 8568      & $0.51\%$\\
$\varepsilon_{0,N_2}$ ($J/mol$)           & $10\,000$     & 7020      & $0.72\%$\\
$\varepsilon_{0,CO_2}$ ($J/mol$)           & $10\,000$     & 7541      & $3.96\%$\\\hline
\end{tabular}
\end{table}

\begin{figure}[!htb]
\centering
\includegraphics[width=12cm]{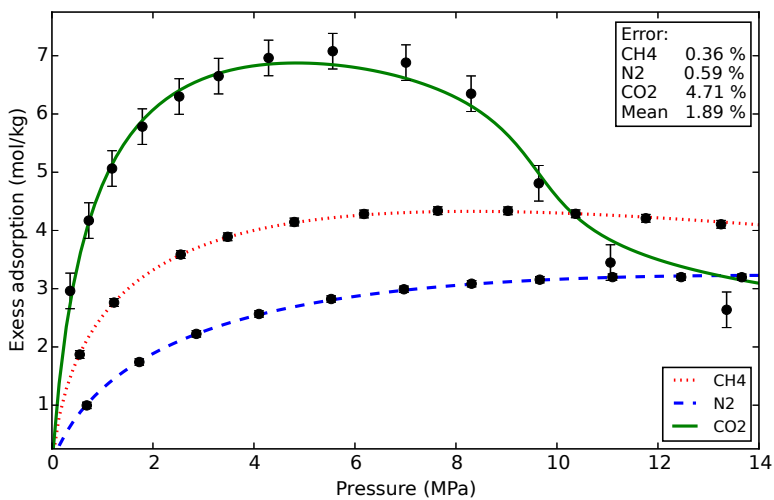}
\caption{\label{Fit_pure_gas}Result of the fit for pure CH$_4$, N$_2$ and CO$_2$ on Calgon F-400\\ at 318.2K for individual values of $z_0$ and $\varepsilon_0$.}
\end{figure}

\begin{figure}[!htb]
\centering
\begin{tikzpicture}[every text node part/.style={align=center}]
\tikzset{code/.style={draw,rectangle,rounded corners=3pt}} 
\tikzset{inputs/.style={draw,rectangle,rounded corners=16pt,thick,fill=blue!10,inner sep=6pt}}
\tikzset{outputs/.style={draw,ellipse,thick,fill=black!5,dotted,fill=green!15}}
\tikzset{radial/.style={very thick,->,>=stealthÕ}}
\node[inputs] (data) at (0,0){Pure gas adsorption data:\\ dataP, dataAd};
\node[inputs] (eos) at (-4,-2){Equation of state:\\ Refprop};
\node[code] (mu) at (-2.25,-2.85){$\mu$};
\node[code] (dmax) at (-2.35,-3.5){$\rho_{Max}$};
\node[code] (z) at (2.75,-3.5){$z$};
\node[inputs] (guess) at (4,-2.5){Initial guesses:\\ $z_0$, $\varepsilon_0$};
\node[inputs] (dra) at (4,-1){Surface potential:\\ DRA};
\node[code] (d) at (0,-2){Convert $P$ to $\rho$:\\ \pythoninline{dB}};
\node[code] (dfunc) at (0,-3.5){Calculate $\rho(z)$:\\ \pythoninline{d_pure()}};
\node[code] (nexcess) at (0,-5){Calculate excess adsorption:\\ \pythoninline{N_ex_pure(dB)}};
\node[code] (model) at (0,-6.5){Compare \pythoninline{N_ex_pure(dB)} with experiment:\\ \pythoninline{pure_fit()}};
\node[code] (min) at (0,-8){Minimize the difference:\\ \pythoninline{N_ex_pure - dataAd}};
\coordinate[shift={(0mm,0mm)}] (n1) at (-5.85,-2);
\coordinate[shift={(0mm,0mm)}] (n2) at (-5.85,-4);
\node[outputs] (z0) at (4.5,-5.5){New $z_0$, $\varepsilon_0$};
\node[outputs] (result) at (-4.25,-8){Final result};
%
%
\draw[->,>=latex,very thick] (data.west)[out=180,in=90] to (eos.north);
\draw[->,>=latex,very thick] (data.south)[out=270,in=90] to (d.north);
\draw[->,>=latex,very thick] (eos.east) to (d.west);
\draw[->,>=latex,very thick] (guess.west)[out=180,in=45] to (dfunc.north east);
\draw[->,>=latex,very thick] (dra.west)[out=180,in=45] to (dfunc.north east);
\draw[->,>=latex,very thick] (eos.south)[out=270,in=180] to (mu.west);
\draw[->,>=latex,very thick] (mu.east)[out=0,in=150] to (dfunc.north west);
\draw[->,>=latex,very thick] (d.south) to (dfunc.north);
\draw[->,>=latex,very thick] (guess.south)[out=270,in=0] to (z.east);
\draw[->,>=latex,very thick] (z.west)[out=180,in=0] to (dfunc.east);
\draw[->,>=latex,very thick] (eos.south)[out=270,in=180] to (dmax.west);
\draw[->,>=latex,very thick] (dmax.east)[out=0,in=180] to (dfunc.west);
\draw[->,>=latex,very thick] (dfunc.south) to (nexcess.north);
\draw[->,>=latex,very thick] (nexcess.south) to (model.north);
\draw[->,>=latex,very thick] (data.west)[out=180,in=90] to (n1)[out=270,in=90] to (n2)[out=270,in=180] to (model.west);
\draw[->,>=latex,very thick] (model.south) to (min.north);
\draw[->,>=latex,very thick,dotted] (min.east)[out=0,in=270] to (z0.south);
\draw[->,>=latex,very thick,dotted] (z0.north)[out=90,in=-45] to (dfunc.-10);
\draw[->,>=latex,very thick,dotted] (nexcess.west)[out=180,in=90] to (result.north);
\end{tikzpicture}
\caption{\label{FlowChart1}Flow chart of the MPTA implementation for pure gases.}
\end{figure}
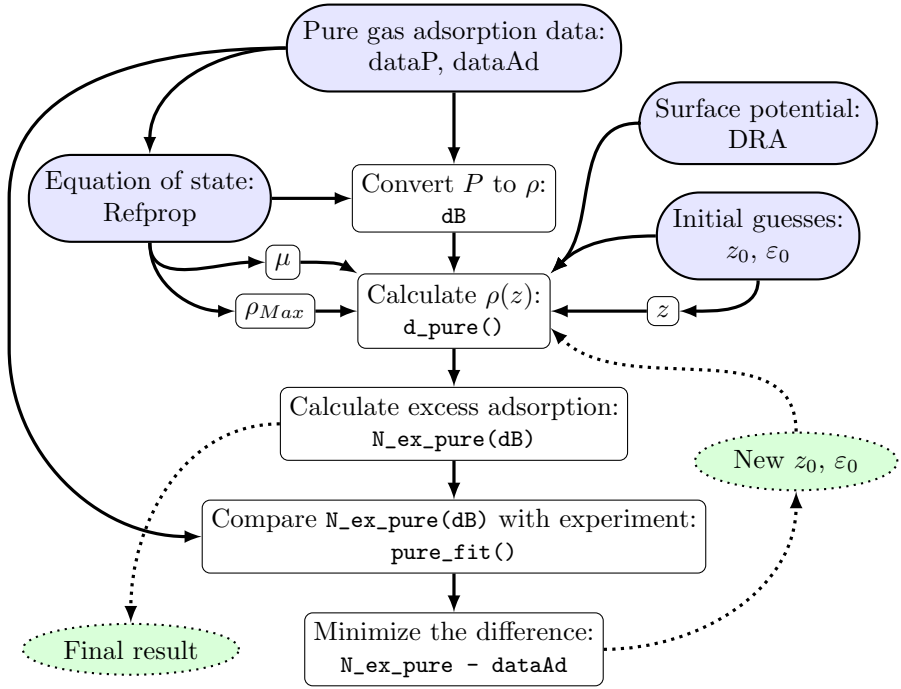

Figure \ref{Fit_pure_gas} give an example for pure CH$_4$, N$_2$ and CO$_2$ (dataset taken from Ref. \cite{Sudibandriyo2003}). The initial values of \pythoninline{z0} and \pythoninline{eps0}, the fitted values and their numerical uncertainty are given in Table \ref{t1}. The mean error between experimental data and the model are given in Table \ref{t3}. Figure \ref{FlowChart1} shows the flow chart of the MPTA implementation for pure gases.

\begin{table}[!htb]
\centering
\caption{Precision of the fits for pure gases on Calgon F-400}\label{t3}
\begin{tabular}{|l|c|c|}
\hline
Gas 		& Mean error (ADDn)	& Experimental uncertainty \cite{Sudibandriyo2003}\\
\hline
\hline
$CH_4$   	& 0.36$\%$	&  1.8$\%$   \\
$N_2$	& 0.59$\%$    	& 2.3$\%$ \\
$CO_2$	& 4.71$\%$    	& 6.4$\%$ \\\hline
Mean	& 1.89$\%$	& 3.5$\%$ \\\hline
\end{tabular}
\end{table}

\subsubsection{Pure gas complications and pitfalls.}

The implementation describes until now work but can present some problems and instabilities even in the case of pure gases. For one thing, the experience had shown that finding the root of

\begin{minipage}{\linewidth}
\begin{python}
0 = chem(T, dB, xB)[0] + DRA(z0, eps0, z) - chem(T, d_z, x_z)[0]
\end{python}
\end{minipage}
in the function \pythoninline{d_pure()}, can strongly depend on the initial guess \pythoninline{d0} for the density. In fact, we remarked that a poor initial guess would be responsible for the majority of instabilities in the code. It is then useful to develop a robust and systematic way of choosing this initial guess. 

\FloatBarrier
First, the function \pythoninline{N_ex_pure()} can be modified in order to use a discrete sum instead of the \pythoninline{scipy.integrate.quad} integration function. Let us divide the interval \pythoninline{(0,z0)} in $N$ subintervals of equal length \pythoninline{delta}. For each subinterval, starting from \pythoninline{z0}, the function \pythoninline{d_pure()} is evaluated, but this time, the result of the preceding subinterval is used as  initial guess for \pythoninline{d0} (of course, for the first subinterval, the bulk density should be a reasonable initial guess). As long as the gas do not go through a phase transition, its density is expected to be a smooth function and then, the density of the gas at \pythoninline{z} should be close to the one at \pythoninline{z+delta}. Explicitly, the function \pythoninline{N_ex_pure()} become

\begin{minipage}{\linewidth}
\begin{python}
def N_ex_pure(T, z0, eps0, dB, xB):
   N  = 300
   delta = z0 / N
   d0 = dB
   integral = 0
   for i in range(0, N):
      d_z = d_pure(z0 - i*delta - delta/2, T, z0, eps0, dB, xB, x_z, d0)
      integral += d_z*delta
      d0 = d_z
   return integral - dB*z0
\end{python}
\end{minipage}
Note that we add the variable \pythoninline{d0} to the function \pythoninline{d_pure()}, which is pass to \pythoninline{fsolve()} to be able to specify the initial guess. Also, we do the sum from \pythoninline{z0} to 0 and not the other way around because the gas density at \pythoninline{z0} is known and given by \pythoninline{dB}. The choice $N=300$ is arbitrary but we obtain a relatively good precision in a reasonable amount of time with this value. 

\bigskip
Now if a phase transition occurs in the adsorbed phase, it can produce a sharp variation in the density such that the smoothness hypothesis no longer holds. In that case, the initial guess for density needs to be adjusted to account for that phase transition. To do that, the density of the gas at dew point and the density of the liquid phase are required. Those density values can be obtained from Refprop. First, we need to ensure that the gas is in the subcritical regime. This can be done by looking at the critical temperature of the gas given by function \pythoninline{refprop.info(icomp=1)['tcrit']}. This function returns the critical temperature of the first gas component (specified with parameter \pythoninline{icomp=1}). With the gas in subcritical regime, the vapor and liquid density at dew point are obtained with

\begin{minipage}{\linewidth}
\begin{python}
refprop._abfl2('TD', T, d, xB)['Dvap']
refprop._abfl2('TD', T, d, xB)['Dliq']
\end{python}
\end{minipage}
In those functions, \pythoninline{'TD'} means that temperature and density are used as input values. Of course, the values of the liquid and vapor density does not depend on the actual fluid density, so we can just put any reasonable number like \pythoninline{d=1} for the density.  Explicitly, we can define the two parameters \pythoninline{d_vap} and \pythoninline{d_liq} as

\begin{minipage}{\linewidth}
\begin{python}
if T < refprop.info(icomp=1)['tcrit']:
   d_vap = refprop._abfl2('TD', T, 1, xB)['Dvap']
   d_liq = refprop._abfl2('TD', T, 1, xB)['Dliq']
else:
   d_vap = d_max(xB)
   d_liq = d_max(xB)
\end{python}
\end{minipage}
Note that if the gas is in the supercritical regime, no phase transition can occur. In that case, we assume that the density vary smoothly and there is no need to adjust the initial guessed density. This is why in that case, \pythoninline{d_vap} and \pythoninline{d_liq} are both set to \pythoninline{d_max}.

Now, the function \pythoninline{d_pure()} can be modified to account for those new parameters. To do that, a condition is added to check if the chemical potential at \pythoninline{z} is greater than the one for the gas with critical density \pythoninline{d_vap}. If so, and if the initial guess \pythoninline{d0} is less than the liquid density \pythoninline{d_liq}, we will set \pythoninline{d0} just over the liquid density (namely \pythoninline{1.1*d_liq}) to help the solver \pythoninline{fsolve()}. Here again, there is the underlying assumption  that the chemical potential is a monotonous increasing function of the density such that $\mu(\rho_{liq})\geq\mu(\rho_{vap})$ because $\rho_{liq}\geq\rho_{vap}$. The \pythoninline{d_pure()} function now become

\begin{minipage}{\linewidth}
\begin{python}
def d_pure(z, T, z0, eps0, dB, xB, x_z, d0):
    def f(d_z):
        y = chem(T, dB, xB)[0] + DRA(z0, eps0, z) - chem(T, d_z, x_z)[0]
        return y
    if chem(T, dB, xB)[0] + DRA(z0, eps0, z) >= chem(T, d_max(xB), xB)[0]:
        return d_max(xB)
    if chem(T, dB, xB)[0] + DRA(z0, eps0, z) > chem(T, d_vap, xB)[0] and d0 < d_liq:
        d0 = 1.1*d_liq
    return scipy.optimize.fsolve(f, d0)[0]
\end{python}
\end{minipage}

\begin{figure}[!htb]
\centering
\includegraphics[width=12cm]{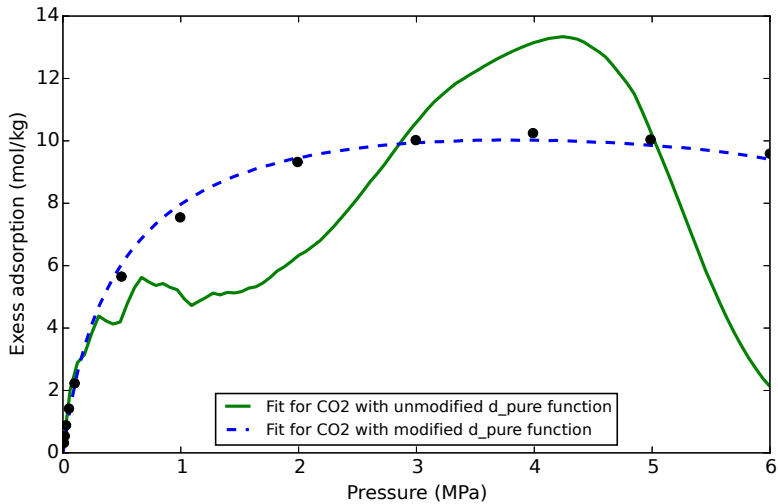}
\caption{\label{Fit-CO2-Norit}Result of the fit on experimental dataset for pure CO$_2$ on Norit-R1 at 298K\\ with and without modifications to d\textunderscore pure() and N\textunderscore ex\textunderscore pure() functions.}
\end{figure}

Those modifications to \pythoninline{d_pure()} and \pythoninline{N_ex_pure()} functions should give a more robust algorithm in case of instabilities and phases transitions. For example, Figure \ref{Fit-CO2-Norit} shows the fit of the model for pure subcritical CO$_2$ on Norit R1 activated carbon at 298\,K, considering original and modified \pythoninline{d_pure()} and \pythoninline{N_ex_pure()} functions (experimental dataset taken from Ref. \cite{dreisbach1999high}). The modified \pythoninline{d_pure()} function exhibit a phase transition of the CO$_2$ from gas to liquid. The fluid phase can be obtain from Refprop using

\begin{minipage}{\linewidth}
\begin{python}
fluid_dict = refprop._abfl2('TD', T, d, x)
print(refprop.getphase(fluid_dict))
\end{python}
\end{minipage}

%

\FloatBarrier
\subsection{Gas mixtures}

Numerical implementation of gas mixtures is a bit more complicated. The MPTA model assumes a common value for $z_0$ to limit the number of adjustable parameters, as discussed by Bj{\o}rner \emph{et~al.} \cite{bjorner2013potential}. This implies that the model must be simultaneously minimized for all pure components isotherms constituting the mixture. This can be achieved readily by building an array, as in function \pythoninline{pure_fit()}, but now containing all the gas components. For example, in a mixture of two gases \pythoninline{A} and \pythoninline{B}, the \pythoninline{difference[i]}, $0\leq i< N_A$ will contain the first component, exactly like in the pure case, while \pythoninline{difference[i]} with $N_A\leq i < N_A+N_B$ will contain the second gas \pythoninline{B}. For simplicity, only binary mixtures will be considered, but generalization to more complex mixtures is straightforward.

It is useful to define the molar fraction of pure gases and mixture in separate arrays. Let us consider a bulk binary mixture of $80\%$ molar fraction for the first gas and $20\%$ molar fraction for the second. We can define the two arrays

\begin{minipage}{\linewidth}
\begin{python}
xB = [0.80 , 0.20]
X  = [[1, 0] , [0, 1]]
\end{python}
\end{minipage}
Here, \pythoninline{xB} is the bulk molar fraction array while \pythoninline{X[i]} represent the molar fraction array of pure gas \pythoninline{A} when $i=0$, and pure gas \pythoninline{B} when $i=1$.
For a ternary mixture, \pythoninline{X} will be given by \pythoninline{X=[[1,0,0],[0,1,0],[0,0,1]]} and so on. 

In the following, we consider a binary mixture composed of $80\%$ CH$_4$ and $20\%$ CO$_2$. The Refprop is initialized accordingly with

\begin{minipage}{\linewidth}
\begin{python}
refprop.setup(u'def', u'METHANE', u'CO2')
\end{python}
\end{minipage}

With those definitions, we can redefine \pythoninline{d_vap} and \pythoninline{d_liq} as

\begin{minipage}{\linewidth}
\begin{python}
d_vap , d_liq = [ ] , [ ]
for i in range(0, len(xB)):
    if T < refprop.info(icomp=i+1)['tcrit']:
        d_vap.append(refprop._abfl2('TD', T, 1, X[i])['Dvap'])
        d_liq.append(refprop._abfl2('TD', T, 1, X[i])['Dliq'])
    else:
        d_vap.append(d_max(X[i]))
        d_liq.append(d_max(X[i]))
\end{python}
\end{minipage}

\noindent
Then, \pythoninline{d_vap[0]} and \pythoninline{d_liq[0]} will returned the values for gas \pythoninline{A} and \pythoninline{d_vap[1]} and \pythoninline{d_liq[1]} will returned the values for gas \pythoninline{B} in the binary mixture.

As explain in Section \ref{Mixtures}, in order to calculate the gas density in the adsorbed phase, a system of equations need to be solved in which the gas molar fraction is one of the variables. A possible problem related to this stems from the fact that when the solver tries to adjust this molar fraction, it can end up with floating point numbers that do not have an exact binary representation. This is known as the \emph{representation error} and then, the program rounds the number and introduce a small error which is usually insignificant. However, Refprop functions have a consistency test that checks if the sum of the molar fractions ends up to precisely 1. Then, the ${\sim10^{-16}}$ error introduced by the rounding can make the consistency test to fail and return errors. One way to circumvent this is to use the package \pythoninline{decimal} that provide arbitrary precision floating point arithmetic. 

Moreover, the argument used to justify the automation of initial guess \pythoninline{d0} for the density now also holds for an initial guess for the gas mixture molar fraction \pythoninline{x0}, as long as the molar fraction vary smoothly. Then, for gas mixtures, the density function now need to receive initial guesses for density and gas molar fractions. Explicitly, the function takes the form

\begin{minipage}{\linewidth}
\begin{python}
def d_mix(z, T, z0, eps0, dB, xB, d0, x0):
    decimal.getcontexr().prec = 100
    def f(u):
        d_z = float(u[0])        
        x_z_A = decimal.Decimal(u[1])    
        x_z_B = decimal.Decimal(1 - x_z_A)
        x_z = [x_z_A , x_z_B]
        y1 = chem(T, dB, xB)[0] + DRA(z0, eps0[0], z) - chem(T, d_z, x_z)[0]
        y2 = chem(T, dB, xB)[1] + DRA(z0, eps0[1], z) - chem(T, d_z, x_z)[1]
        return [y1 , y2]
    if chem(T, dB, xB)[0] + DRA(z0, eps0[0], z) >= chem(T, d_max(X[0]), xB)[0] or \
        chem(T, dB, xB)[1] + DRA(z0, eps0[1], z) >= chem(T, d_max(X[1]), xB)[1]:
        return [d_max(xB) , x0[0]]
    if chem(T, dB, xB)[0] + DRA(z0, eps0[0], z) > chem(T, d_vap[0], xB)[0] and \
        d0 < d_liq[0]:
        d0 = 1.1 * d_liq[0]
    if chem(T, dB, xB)[1] + DRA(z0, eps0[1], z) > chem(T, d_vap[1], xB)[1] and \
        d0 < d_liq[1]:
        d0 = 1.1 * d_liq[1]
    return = scipy.optimize.fsolve(f, [d0 , x0[0]])
\end{python}
\end{minipage}
The second line set the precision for the \pythoninline{Decimal()} function. With this precision, the representation error is $\sim10^{-51}$ which is small enough  for the consistency test of the Refprop functions to succeed. Most of the properties and parameters referring to a pure gas are now described by arrays of two values, one for each gas component.  Then, \pythoninline{chem(...)[i]}, \pythoninline{eps0[i]}, \pythoninline{d_vap[i]} and \pythoninline{d_liq[i]} refer to the component \pythoninline{A} for \pythoninline{i=0} and component \pythoninline{B} for \pythoninline{i=1}. The variables \pythoninline{dB}, \pythoninline{xB} refer to the bulk density and molar fractions while \pythoninline{d_z} and \pythoninline{x_z} are the unknowns and refer to the density and the molar fraction of the adsorbed phase at distance \pythoninline{z} from the adsorbent surface. The function \pythoninline{fsolve()} now have to solve a system of two equations. To proceed, we define a function \pythoninline{f(u)}, with \pythoninline{u} an array containing the gas density \pythoninline{d_z} and the molar fraction of the first component \pythoninline{x_z_A} to be found. This function returns an array containing the two equations to be solved. The solver now requires initial guesses in an array format to match the system of equation, which is given by \pythoninline{[d0,x0[0]]}. Remark that for a mixture of two gases, we only used the molar fraction of the first component and get the molar fraction of the second component by \pythoninline{1 - x0[0]}. In the function \pythoninline{f(u)}, the \pythoninline{Decimal()} function is used to redefine the molar fraction to \pythoninline{x_z_A} and \pythoninline{x_z_B} such that \pythoninline{1-x_z_A-x_z_B}$\sim10^{-51}$. Finally, the function \pythoninline{d_mix()} return an array composed of the gas density and the molar fraction of the first component. 

This \pythoninline{d_mix()} function induced modifications to the \pythoninline{N_ex_pure} function that becomes

\begin{minipage}{\linewidth}
\begin{python}
def N_ex_mix(T, z0, eps0, dB, xB):
    N  = 300
    delta = z0 / N
    d0 = dB
    x0 = xB 
    integral_A , integral_B = 0 , 0
    for i in range(0, N):
        [d_z, x_z_A] = d_mix(z0 - i*delta - delta/2, T, z0, eps0, dB, xB, d0, x0)
        integral_A += d_z*x_z_A*delta
        integral_B += d_z*(1 - x_z_A)*delta
        d0 = d_z
        x0 = [x_z_A , 1 - x_z_A]
    return [integral_A - dB*xB[0]*z0 , integral_B - dB*xB[1]*z0]
\end{python}
\end{minipage}
In this function, \pythoninline{d_mix()} return the two variables \pythoninline{d_z} and \pythoninline{x_z_A}. Variables \pythoninline{integral_A} and \pythoninline{integral_B} represent the excess adsorption for each component in the mixture. As we can see, the terms \pythoninline{d_z*x_z_A*delta} and \pythoninline{d_z*(1-x_z_A)*delta} give the absolute adsorption for component \pythoninline{A} and \pythoninline{B} respectively, in the interval \pythoninline{delta}.  The initial guesses \pythoninline{d0} and \pythoninline{x0}, for density and molar fraction, are updated with newly calculated values \pythoninline{d_z} and \pythoninline{[x_z_A ,1-x_z_A]}. Finally, the (Gibbs) excess adsorption for each component are given by the absolute adsorption minus the term \pythoninline{dB*xB[0]*z0} or \pythoninline{dB*xB[1]*z0}.

The \pythoninline{mix_fit()} function for mixtures take the form

\begin{minipage}{\linewidth}
\begin{python}
def mix_fit(params):
    value = params.valuesdict()
    z0 = value['z0']
    eps0_A = value['eps0_A']
    eps0_B = value['eps0_B']
    difference = [ ]
    for i in range(0, len(dataD_A)):
        difference.append(N_ex_pure(T, z0, eps0_A, dataD_A[i], X[0]) - dataAd_A[i])
    for i in range(0, len(dataD_B)):
        difference.append(N_ex_pure(T, z0, eps0_B, dataD_B[i], X[1]) - dataAd_B[i])
    return difference
\end{python}
\end{minipage}
Of course, \pythoninline{N_ex_pure()} function is used here and not \pythoninline{N_ex_mix()} because the purpose of \pythoninline{mix_fit()} is to optimized the fit of pure gas \pythoninline{A} and \pythoninline{B} simultaneously with an unique \pythoninline{z0}. Finally, the fit is executed with

\begin{minipage}{\linewidth}
\begin{python}
params = lmfit.Parameters()
params.add('z0', value=1, min=0, vary=True)
params.add('eps0_A', value=10000, min=0, vary=True)
params.add('eps0_B', value=10000, min=0, vary=True)
result = lmfit.minimize(mix_fit, params)
print(lmfit.fit_report(result))
\end{python}
\end{minipage}

\begin{figure}[!htb]
\centering
\includegraphics[width=12cm]{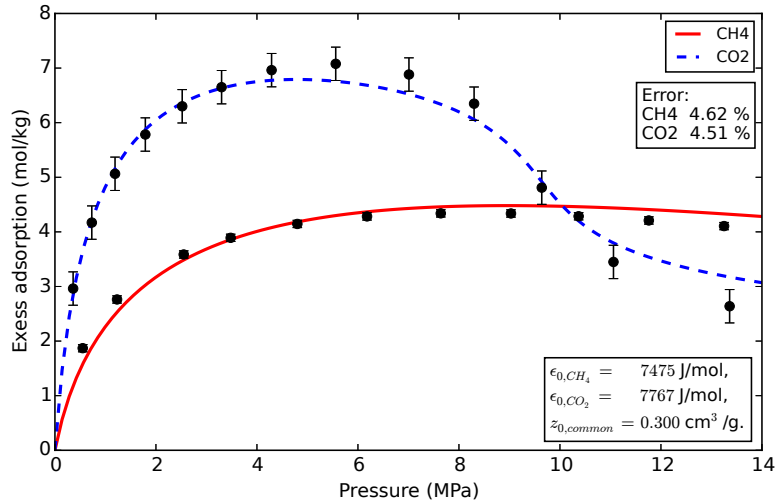}
\caption{\label{Fit_CH4_CO2}Result of the fit for pure CH$_4$ and CO$_2$ on Calgon F-400 at 318.2K.}
\end{figure}

For a CH$_4$~/~CO$_2$ mixture on Calgon F-400 at 318.2K \cite{Sudibandriyo2003}, our fitted parameters are given in Table \ref{params_mix}. Figure \ref{Fit_CH4_CO2} represents the fit for pure gases with a common \pythoninline{z0}. The mean error between experimental data pure isotherms and the fitted MPTA model are $4.62\%~/~4.51\%$ respectively. 

\begin{table}[!htb]
\centering
\caption{Values of the parameters for CH$_4$~/~CO$_2$ mixture\\ on Calgon fitted from pure gas isotherms.}\label{params_mix}
\begin{tabular}{|l|c|c|c|c|}
\hline
Parameter & Initial value & Fitted value & Fit error $\%$\\
\hline
$z_{0,common}$ ($cm^3/g$)                    & 1             & 0.300        & $1.79$ \\
$\varepsilon_{0,CH_4}$ ($J/mol$)     & 10\,000         & 7475         & $3.08$\\
$\varepsilon_{0,CO_2}$ ($J/mol$)        & 10\,000         & 7767        & $3.45$\\\hline
\end{tabular}
\end{table}

\begin{figure}[!htb]
\centering
\includegraphics[width=12cm]{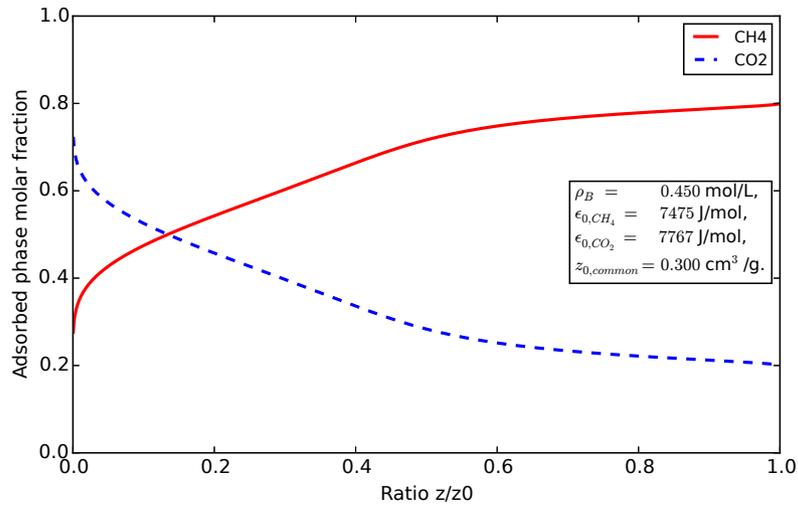}
\caption{\label{MolarFraction-CH4-CO2}Adsorbed phase molar fraction of each component for a\\ $80\%$~CH$_4$~/~$20\%$~CO$_2$ mixture on Calgon F-400 at 318.2K.}
\end{figure}

\begin{figure}[!htb]
\centering
\includegraphics[width=12cm]{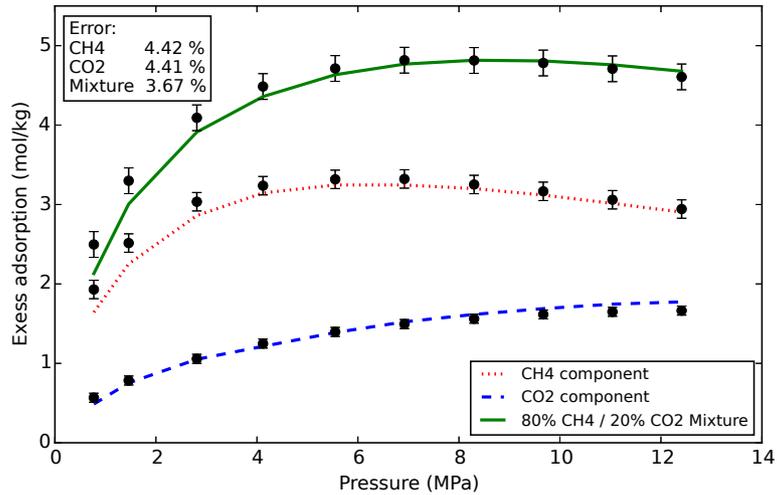}
\caption{\label{CH4-CO2-mixture}Excess adsorption for a $80\%$~CH$_4$~/~$20\%$~CO$_2$ mixture on Calgon F-400 at 318.2K.}
\end{figure}

Figure \ref{MolarFraction-CH4-CO2}, presents the molar fraction of each gas component near the surface of the adsorbent.
Finally, in Figure \ref{CH4-CO2-mixture}, we present the (Gibbs) excess adsorption for a $80\%$~CH$_4$~/~$20\%$~CO$_2$ mixture. The figure shows the excess adsorption of each component and the total excess adsorption for the mixture. 
The error bars represent the expected experimental uncertainties on the measurements given in Ref. \cite{Sudibandriyo2003}. The mean error between experimental data and the model are presented in table \ref{t5}. It is worth noting that the mean error of the MPTA model is of the same order as the experimental uncertainty on datasets.

To give an idea of the importance of bulk phase molar fraction values at equilibrium, we can recalculate the excess adsorption of this particular mixture using only the feeding molar composition, which differs from the measured molar fraction by at most $10\%$. The resulting mean error between experimental and calculated values increases from $4.42\%$, $4.41\%$ and $3.67\%$ to  $25\%$, $71\%$ and $4.15\%$ for the CH$_4$ component, the CO$_2$ component and the mixture, respectively.

\begin{table}[!htb]
\centering
\caption{Mean error between experimental data and the MPTA model for\\ a $80\%$~CH$_4$~/~$20\%$~CO$_2$ mixture on Calgon.}\label{t5}
\begin{tabular}{|l|c|c|}
\hline
& Model mean error (ADDn)  & Experimental uncertainty \cite{Sudibandriyo2003} \\\hline
\hline
CH$_4$     & 4.42$\%$    & 4.0$\%$ \\
CO$_2$     & 4.41$\%$     & 4.3$\%$ \\\hline
Mixture    & 3.67$\%$     & 3.9$\%$ \\\hline
\end{tabular}
\end{table}
%

\FloatBarrier
\section{Conclusion}

We presented a detailed numerical implementation of the MPTA adsorption model, covering pure gases and binaries mixtures. From our treatment of the binary mixture case, generalization to more complex mixtures is straightforward. This implementation includes some of the strategies used to obtain a robust model even in case of fast variation of fluid densities, like phases transitions.  For all the mixtures in Ref. \cite{Sudibandriyo2003}, the mean error between predictions of the MPTA model for mixture (Gibbs) excess adsorption and experimental data vary from $1.44\%$ ($20\%$ CH$_4\,$/$\,80\%$ CO$_2$ mixture) to $5.09\%$ ($20\%$ N$_2\,$/$\,80\%$ CO$_2$ mixture). Meanwhile, the experimental uncertainties for mixture adsorption vary from $2.4\%$ ($80\%$ CH$_4\,$/$\,20\%$ N$_2$ and $40\%$ CH$_4\,$/$\,60\%$ N$_2$ mixtures) to $4.1\%$ ($80\%$ N$_2\,$/$\,20\%$ CO$_2$ mixture). 

For a complete version of the code, the interested reader can consult:\\ 
\indent
\url{https://github.com/RaphaelGervaisLavoie/MPTA}

\section*{Acknowledgements}

The authors would like to thanks Ege Dundar for his help and suggestion about the implementation and also the Natural Sciences and Engineering Research Council of Canada and the Savannah River National Laboratory for financial support.

\bibliographystyle{ieeetr}
\bibliography{biblio.bib}

\end{document}